\documentclass[lettersize, journal]{IEEEtran}

\usepackage{cite}
\usepackage{amsmath,amssymb,amsfonts}
\usepackage{textcomp}
\usepackage{tikz}
\usepackage{amsmath}
\usepackage{enumitem}
\usepackage{diagbox}
\usepackage{array,multirow,graphicx,makecell}
\usepackage{listings,amsfonts}
\usepackage{bigstrut}
\usepackage{threeparttable}
\usepackage{booktabs}
\usepackage{bm}
\usepackage{fixltx2e}
\usepackage{hyperref}
\usepackage{framed}
\usepackage{booktabs,caption}
\usepackage{subcaption}
\usepackage{amsmath,xcolor,pifont}
\usepackage{xspace}
\usepackage{float}
\usepackage{color}
\usepackage{ifthen}
\usepackage[T1]{fontenc}
\usepackage{enumerate}
\usepackage{balance}
\usepackage{tikz}
\usepackage{calc}
\usepackage{url}
\usepackage{hyperref,endnotes}
\usepackage[misc]{ifsym}
\usepackage{amsmath}
\usepackage{mathtools}
\usepackage{algorithm}
\usepackage[noend]{algorithmic}
\usepackage[affil-it]{authblk}

\floatname{algorithm}{Algorithm}

\begin{document}\sloppy

\title{{\tool}: A Price Manipulation Detection Service in DeFi using Graph Neural Networks}

\author[1]{Dabao Wang}
\author[1]{Bang Wu}
\author[2]{Xingliang Yuan}
\author[3]{Lei Wu}
\author[3]{Yajin Zhou}
\author[4]{Helei Cui}

\affil[1]{Monash University}
\affil[2]{The University of Melbourne}
\affil[3]{Zhejiang University}
\affil[4]{Northwestern Polytechnical University}

\newcommand{\mynote}[2]{
    \fbox{\bfseries\sffamily\scriptsize#1}
    {\small$\blacktriangleright$\textsf{\emph{#2}}$\blacktriangleleft$}}
\newcommand*{\prompt}[1]{\textsf{\textbf{#1}}}
\newcommand{\tab}{\hspace*{1em}}
\newcommand{\code}[1]{{\fontfamily{cmtt}\fontseries{m}\fontshape{n}\selectfont\small{#1}}}
\newcommand{\fixme}[1]{\textcolor{red}{#1}}
\newcommand{\new}[1]{\textcolor{blue}{#1}}

\newcommand{\term}[1]{\( \mathcal{#1} \)\xspace}
\newcommand{\equa}[1]{\( #1 \)\xspace}
\newcommand{\dapp}{DApp\xspace}
\newcommand{\defi}{DeFi\xspace}
\newcommand{\pma}{PMA\xspace}
\newcommand{\tool}{{\it DeFiGuard}\xspace}
\newcommand{\cmark}{\ding{51}}%
\newcommand{\xmark}{\ding{55}}%

\newcommand*{\circled}[1]{\tikz[baseline=(char.base)]{
    \raisebox{0mm}{\node[shape=circle,draw,inner sep=0.5pt] (char) {\footnotesize #1};}}}

\newcommand*{\ellipsed}[1]{\tikz[baseline=(char.base)]{
    \raisebox{0mm}{\node[shape=ellipse,draw,inner sep=0.5pt] (char) {\footnotesize #1};}}}

\newcommand*{\encircled}[1]{\tikz[baseline=(char.base)]{
	\raisebox{0mm}{\node[shape=circle,draw,inner sep=0.5pt, fill=black, text=white] (char) {\footnotesize #1};}}}

\maketitle

\begin{abstract}
The prosperity of Decentralized Finance (DeFi) unveils underlying risks, with reported losses surpassing 3.2 billion USD between 2018 and 2022 due to vulnerabilities in Decentralized Applications (DApps). 
One significant threat is the Price Manipulation Attack (PMA) that alters asset prices during transaction execution.
As a result, PMA accounts for over 50 million USD in losses.
To address the urgent need for efficient PMA detection, this paper introduces a novel detection service, \tool{}, using Graph Neural Networks (GNNs). 
In this paper, we propose cash flow graphs with four distinct features, which capture the trading behaviors from transactions.
Moreover, \tool{} integrates transaction parsing, graph construction, model training, and PMA detection. 
Evaluations on a dataset of 208 PMA and 2,080 non-PMA transactions show that \tool{} with GNN models outperforms the baseline in Accuracy, TPR, FPR, and AUC-ROC. 
The results of ablation studies suggest suggest that the combination of the four proposed node features enhances \tool{}'s efficacy.
Moreover, \tool{} classifies transactions within 0.892 to 5.317 seconds, which provides sufficient time for the victims ({\dapp}s and users) to take action to rescue their vulnerable funds.
In conclusion, this research offers a significant step towards safeguarding the DeFi landscape from PMAs using GNNs.

\end{abstract}

\section{Introduction}
\label{sec:introduction}
The surge of Decentralized Finance (DeFi) since 2019 has attracted a significant influx of capital into the blockchain ecosystem. 
On the one hand, various entities are constructing decentralized applications ({\dapp}s) to offer financial services within the \defi ecosystem. 
On the other hand, users are increasingly drawn to engage in interacting with these {\dapp}s to gain profits.
However, the prosperity of \defi not only introduces the chance of making financial gains but also brings the underlying risks of losing assets~\cite{qin2022quantifying, qin2023blockchain, daian2020flash, zhou2020highfrequency, wu2021defiranger}. 
The research~\cite{zhou2023sok} reveals the alarming fact that the code and logic vulnerabilities in {\dapp}s resulted in a staggering loss exceeding 3.2 billion USD between April 2018 and April 2022.

The price manipulation attack (PMA), one of the infamous attacks within the {\defi} realm, has attracted considerable attention and resulted in financial losses exceeding 200 million USD~\cite{slowmist, defiRekt}. 
Specifically, the {\pma} refers to the malicious exploitation of vulnerabilities in smart contracts to manipulate asset prices by tampering with the price oracles or altering the trading ratio in exchanges' liquidity pools. 
Initiating a {\pma} typically necessitates multiple interactions across a range of {\dapp}s. 
Compounding the challenge, astute attackers frequently deploy opaque smart contracts to obscure the execution logic, leading to unexpected state alterations, such as inflating or deflating the asset price.
%


Existing research~\cite{wu2021defiranger} extracts semantic knowledge from a collection of {\dapp}s to establish two expert-defined patterns, which identify two kinds of {\pma}s: direct and indirect {\pma}s.
However, this reliance on predefined patterns may impede the comprehensive detection of {\pma}s. 
As these patterns are tied to the semantic understanding of a variety of {\dapp}s, their adaptability and scalability are limited due to the rapid increase in the number of {\dapp}s.
Therefore, there is an urgent need to design a highly adaptive methodology capable of detecting an extensive array of {\pma}s.

Compared to rule-based methodologies, the learning-based approach can offer adaptability in the field of vulnerability detection~\cite{zhuang2021smart}.
In this study, we construct the cash flow graph for the transaction and conduct a graph representation learning using Graph Neural Networks (GNNs) for {\pma} detection based on two insights.
First, after examining various {\pma}s, we discover that the cash flow graph provides a comprehensive representation of trading information for a transaction.
Specifically, the account can be represented as a node and the asset transfer can be represented as an edge in the graph.
Second, recent advancements in the field have seen the success of GNNs for handling graph-structured data in the field of cybersecurity~\cite{wang2019mcne, fan2019graph, bian2020rumor}.
Given these insights, we believe in the potential of training a GNN model to capture the nuances of the cash flow graphs for {\pma} detection.

In this study, we introduce \tool{}, a detection service for {\pma} detection using GNNs. 
To ensure swift updates in response to evolving detection methodologies and GNN algorithms, we structured our service into three components: {\it Transaction Parser}, {\it Graph Builder}, and {\it Graph Classifier}.
1) The {\it Transaction Parser} collects the raw transactions from EVM-based blockchains and retrieves transfer-related call traces and event logs.
2) The {\it Graph Builder} constructs cash flow graphs with four distinct node features based on the call traces and event logs received from the {\it Transaction Parser}.
3) The {\it Graph Classifier}  builds a GNN model tailored for graph representation learning and employs a pre-trained GNN model for {\pma} detection.
The decomposing nature of \tool{} provides a flexible foundation for future improvements, such as updating the {\it Graph Classifier} with cutting-edge GNN algorithms.

In our evaluation, we assessed the performance of \tool{} using a dataset of 208 PMA and 2,080 non-PMA transactions.
The overall performance reveals that when paired with GNN models, \tool{} outperformed the baseline MLP model across metrics such as {\it Accuracy}, {\it TPR}, {\it FPR}, and {\it AUC-ROC} value. 
For instance, when paired with the GraphSage model, \tool{} achieves a peak performance of 93.25\% {\it Accuracy}, 91.67\% {\it TPR}, 6.67\% {\it FPR}, and 96.18\% {\it AUC-ROC}.
Moreover, the hyperparameter tuning process shows that training with only 100 PMA and non-PMA transactions across 100 epochs yields strong performance. This further underscores the effectiveness of the cash flow graph in capturing transaction trading behavior.
In the ablation studies, we evaluate the impact of the four proposed node features: {\it node type}, {\it transfer frequency}, {\it transfer diversity}, and {\it profit score} on classification performance. 
The results suggest that the classification performance for {\pma} detection stems from the collective impact of these four node features, with no single feature predominating.
Regarding practicality, we evaluated the time consumed by \tool{} for {\pma} detection. 
The overall for classifying a single transaction ranged from 0.892 to 5.317 seconds, which is faster than the time required to generate a new block on Ethereum. 
This implies that, with a well-connected node, the potential victims, including {\dapp}s and users, can have sufficient time to secure their vulnerable assets after receiving the classification result on provided transactions.

To sum up, this work makes the following contributions:
\begin{itemize}
    \item This paper introduces a novel detection service, \tool{}, designed to detect PMAs using GNNs. 
    \tool{} is a comprehensive solution that seamlessly integrates transaction parsing, graph construction, model training, and PMA detection.
    \item We present a unique approach to encapsulate trading behavior by translating raw transactions into cash flow graphs, enriched with features: {\it node type}, {\it transfer frequency}, {\it transfer diversity}, and {\it profit score}. 
    Our findings indicate that the constructed cash flow graph with these four features significantly facilitates PMA detection.
    Moreover, we will release the graph-based dataset after the publication.
    \item We evaluated \tool{}'s performance in PMA detection using various metrics. 
    Empowered by selected GNN models, \tool{} outperforms the baseline MLP model across metrics such as Accuracy, TPR, FPR, and AUC-ROC. 
    An ablation study further confirms the importance of each feature to \tool{}'s effectiveness. 
    Notably, \tool{} can classify a transaction in 0.892 to 5.317 seconds, faster than Ethereum's block generation time of approximately 13 seconds.

\end{itemize}

\section{Background}
\label{sec:background}

\subsection{The Primitives of EVM-based Blockchain}
\label{subsec:blockchain}
\noindent \textbf{EVM-based blockchain.} Blockchain technology, first introduced in 2008 by Satoshi Nakamoto~\cite{nakamoto2008bitcoin}, revolutionized digital currencies and enabled secure, decentralized financial transactions. 
The primary purpose of blockchain technology is to create a trustless, transparent, and tamper-proof digital ledger, recording transactions in a peer-to-peer (P2P) network.
The participants in the P2P network can validate and store transaction data, thereby eliminating the need for centralized intermediaries such as banks. 
Transactions in the network are grouped into blocks and cryptographically secured using a consensus algorithm called Proof-of-Work (PoW). 
As a result, the blockchain provides an immutable record of all transactions since the inception of the network, ensuring data integrity and security.
In 2014, Ethereum~\cite{wood2014ethereum} was introduced to enable the execution of programmable agreements (i.e., smart contracts written in a Turing-completed language), allowing developers to build decentralized applications ({\dapp}) that can automate complex processes.
Specifically, the Ethereum blockchain utilizes the Ethereum Virtual Machine (EVM) as a runtime environment for executing smart contracts.
Ethereum and Binance Smart Chain~\footnote{A EVM-compatible blockchain network launched by Binance in September 2020.} are two prominent examples of EVM-based blockchains.
%

\noindent \textbf{Accounts.}
%
%
In EVM-based blockchain systems, an account refers to a digital entity representing either a user or a smart contract. There are two types of accounts: Externally Owned Accounts (EOAs) and Contract Accounts (CAs). 
EOAs are controlled by individuals who own the corresponding private key, while CAs are controlled by smart contracts, which are snippets of JavaScript-like code.
To create a CA, users must initiate a signed transaction that deploys their smart contract.
This process generates a new address on the blockchain, which can handle and manage digital assets, perform complex business logic, and execute other smart contracts.

\noindent \textbf{Transaction Call Trace.}
The transaction call trace in an EVM-based blockchain provides a detailed account of the execution flow of a transaction.
Essentially, it is a step-by-step outline of how a transaction is processed, illustrating every action taken from the transaction's initiation to its conclusion.
The call trace captures the intricate interactions within and between smart contracts.
Within a call trace, vital information is presented, including the caller account (initiator of a call), the callee account, the invocation call data, and the associated value (any native asset sent in the call).


\noindent \textbf{Transaction Event Log.}
Transaction event logs in EVM-based blockchain systems are immutable records produced during smart contract execution.
They offer transparency by logging pre-defined events in smart contracts.
Analyzing these logs provides insights into contract interactions and outcomes. 
In this paper, we focus on the \code{Transfer} event in ERC20 standard token contracts, which log token transfers between accounts. 
For example, the parameters of the event \code{Transfer(address from, address to, uint256 value)} typically include the sender's address (\code{from}), the recipient's address (\code{to}), and the number of tokens being transferred (\code{value}).
Moreover, the event emitter for this Transfer event is usually the ERC20 token contract address.
The \code{Transfer} event offers a transparent record of all ERC20 token movements, ensuring traceability within the ERC20 ecosystem.

\subsection{Graph Neural Networks}
\label{subsec:background_gnn}
Graph neural networks (GNNs) are a type of powerful machine learning method designed to work with graph-structured data. 
They have made significant advances across various applications, such as smart contract vulnerabilities detection~\cite{liu2021smart}\cite{zhuang2021smart}\cite{liu2021combining} and other security analysis~\cite{yang2022wtagraph}\cite{kingeuler}. 
Unlike other deep neural networks operating on Euclidean data (e.g., analyzing images via convolution neural networks), GNNs can gather graph structure information by iteratively aggregating each node's information among their connections in a graph, which makes them an effective and useful tool for graph data exploration. 
Specifically, they employ message-passing algorithms to propagate information along the edges of a graph and update the node features. 
The updated node features can then be used for various downstream graph analysis tasks, such as node classification, link prediction, and graph classification. 
%
A typical GNN model will iteratively update each node's representation (also known as node embedding) as:

\begin{equation}
\label{eq:GNN}
\begin{aligned}
\mathbf{m}_{v}^{(t)}&=\sum_{u\in\mathcal{N}(v)}M_{t}(\mathbf{h}_u^{(t-1)},\mathbf{h}_v^{(t-1)}, \mathbf{e}_{uv}), \\
\mathbf{h}_v^{(t)} &= U_t(\mathbf{h}_v^{(t-1)}, \mathbf{m}_v^{(t)}),
\end{aligned}
\end{equation} 

\noindent where $\mathbf{h}_v^{(t)}$ indicates the embedding of node $v$ at layer $t \in \{1,\ldots,T\}$, and $\mathcal{N}(v)$ denotes the set of neighbours of $v$ in graph $G$. $M_t(\cdot,\cdot,\cdot)$ and $U_t(\cdot,\cdot)$ are the message function and the embedding updating function at layer $t$, respectively. 
Once the embeddings for each node in a graph are produced, the representation of the entire graph can be generated by combining them (e.g., calculating the average value), which can then be used for downstream tasks (e.g., predicting the label of a graph).

\section{Problem Definition}
\label{sec:problem}

\subsection{Problem}
Given the proven efficacy of GNNs in graph learning tasks, we propose to depict transactions as graph data for convenient analysis and formulate cash flow graphs, denoted as \term{G}.
Specifically, we use edges \term{E} to represent directed asset transfers, and nodes \term{V} to symbolize accounts sending or receiving assets.
Therefore, the PMA detection task, which determines whether a transaction is executing a PMA for illicit financial gain, can be conceptualized as a task of GNN-based graph representation learning followed by classification.
Formally, given a cash flow graph \term{G}, the objective is to build a graph classification model $\mathcal{F}$, which outputs a prediction $\mathcal{P}$.
The prediction determines if transactions are PMAs ($\mathcal{P}=1$) or non-PMAs ($\mathcal{P}=0$), as illustrated in Equation~\ref{eq:problem}.

\begin{equation}
\label{eq:problem}
\begin{aligned}
    \mathcal{P} \rightarrow \mathcal{F}(\mathcal{G}), where\ \mathcal{P}=\ 0\ or\ 1
\end{aligned}
\end{equation}

\subsection{Requirements}
To accomplish the aforementioned task within the context of the DeFi domain, we enumerate several fundamental requirements.

\noindent \textbf{Effectiveness.}
The proposed methodology should reliably detect an extensive spectrum of PMAs, with a strong performance on key metrics like \textit{Accuracy}, \textit{True Positive Rate}, \textit{False Positive Rate}, and \textit{AUC-ROC}.

\noindent \textbf{Generability.}
The proposed methodology must learn and understand diverse behaviors with {\dapp}s, capture attacks from a limited dataset, and identify unknown attacks using various GNN algorithms.

\noindent \textbf{Automation.}
System automation is crucial, spanning from transaction parsing to classification, ensuring seamless PMA detection in real time.

\noindent \textbf{Adaptability.}
The proposed methodology should prioritize adaptability and quick updates to stay current with emerging detection techniques, including new transaction analysis methods and advanced GNN algorithms.
\section{Design}
\label{sec:design}

\begin{figure*}[ht]
    \centering
    \includegraphics[width=0.9\linewidth]{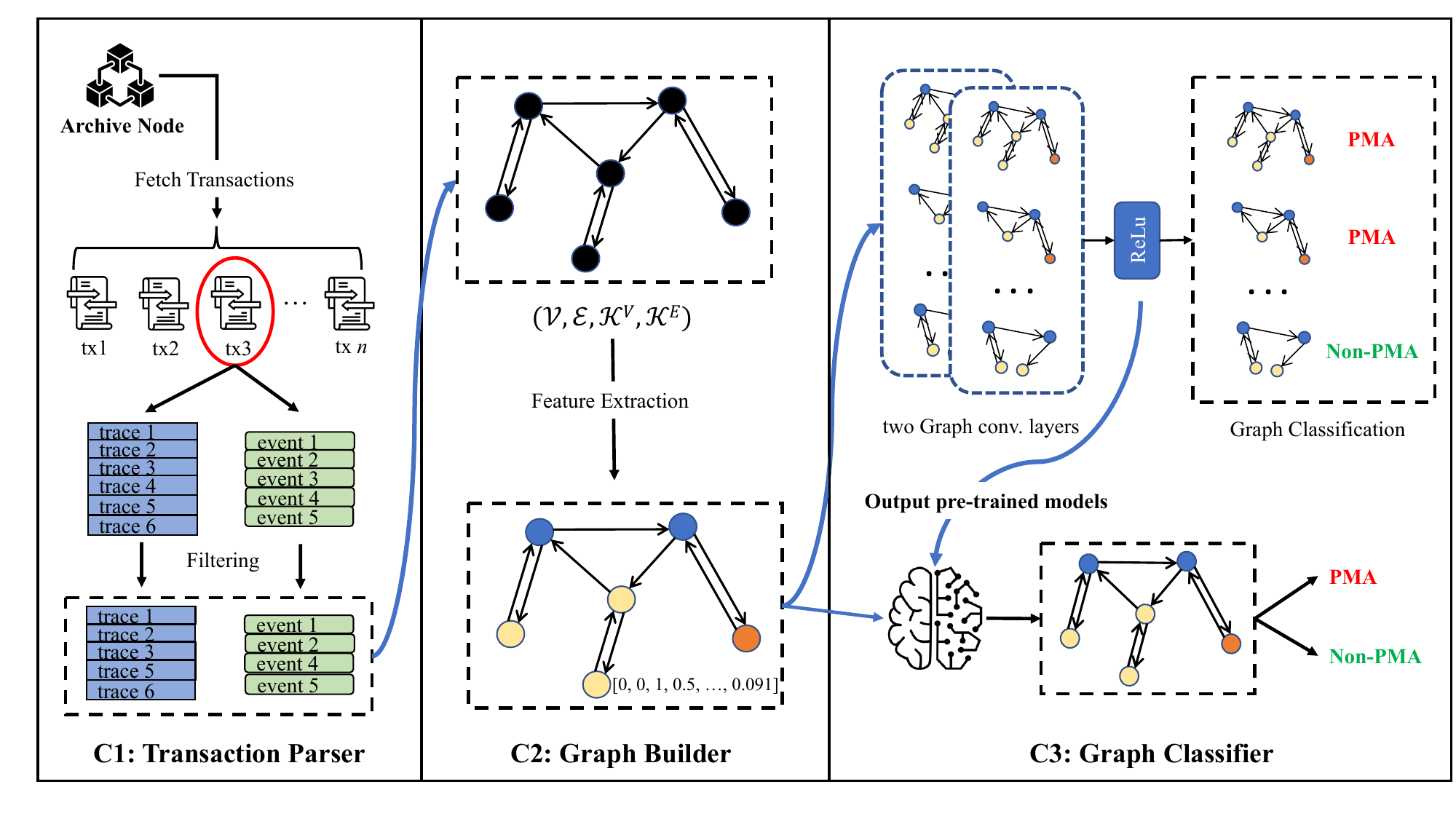}
    \caption{\bf The high-level architecture of {\tool}.} 
    \label{fig:design_overview}
\end{figure*}

In this section, we first identify two challenges in our task.
Then, we elaborate on the design rationale and the high-level architecture of our PMA detection framework {\tool}.

\subsection{Challenges}
\label{subsec:challenges}
After formulating our problem as a task of learning graph representation and classification in a cash flow graph, we identified the following two main challenges.

\noindent {\bf Detecting an extensive spectrum of PMAs.}
Analyzing a diverse array of PMAs poses considerable difficulty when employing pattern-based detection techniques due to the evolving nature of the \defi ecosystem. 
Prior work, e.g., \textit{DeFiRanger}, proposes two patterns focusing on detecting direct and indirect PMAs, it strongly relies on the semantic knowledge extracted from {\dapp}s, which requires a significant effort to update their patterns to detect new PMAs.
To effectively overcome the evolving nature of such attacks, it becomes imperative for the pattern design to exhibit robust adaptability and versatility.
Consequently, the endeavor of crafting and revising these patterns demands substantial effort and resources.

\noindent {\bf Building an efficient and adaptive framework for PMA detection.}
Detecting {\pma}s in the EVM-based blockchains presents a challenge, particularly with regard to the temporal implications. 
Notably, the Ethereum network exhibits a mining duration of approximately 12 to 15 seconds for each new block.
Consequently, if the task of detection and alerting fails to occur (i.e., prior to the subsequent block), the attack's impact may intensify and disseminate more extensively.

\subsection{Cash Flow Graph}
\label{subsec:cfg}

As aforementioned, it is challenging to identify a broader spectrum of PMAs with fixed patterns due to the evolving nature of PMAs in the {\defi} ecosystem.
In this paper, we introduce the concept of the cash flow graph, which encapsulates comprehensive information regarding asset transfers, making it a robust representation of trading behaviors. 
Specifically, the cash flow graph inherently embodies an abundance of information (such as {\it transfer frequency}, {\it transfer diversity}, and {\it profit score}) that can unveil malicious trading activities.
Moreover, leveraging the cash flow graph with GNN techniques allows for a dynamic approach rather than adhering to static rule patterns.
When training on the cash flow graph, the GNN model can discern and learn the intricacies of various PMAs.
This facilitates the detection of a broader spectrum of PMAs, making it a more adaptive and robust solution.

To facilitate a deeper understanding of the proposed cash flow graph, we elucidate a list of pertinent terminologies:

\begin{description}[leftmargin=10pt]
    \item[Node \term{V}:]
    The node, denoted by \term{V}, represents a set of accounts sending or receiving assets.
    In this study, we classify nodes based on their account type and transparency.
    \item[Edges \term{E}:]
    The edge in the graph represents the asset transfer in terms of the native asset and the ERC20 standard asset.
    Specifically, \(\mathcal{E} \subseteq  (\mathcal{V}\) X \(\mathcal{V})\) is a set of directed edges.
    \item[Graph Metadata \term{K}:]
    The graph metadata, denoted by \term{K}, comprises \(\mathcal{K}^{V} \) and \(\mathcal{K}^{E}\).
    It is extracted during the construction of the cash flow graph.  
    Specifically, \(\mathcal{K}^{V}\) represents the metadata of nodes, which includes the account address, while \(\mathcal{K}^{E}\) signifies the metadata of edges, encompassing both asset address and asset amount.
    \item[Node Feature \term{X}:]
    Node feature, denoted by \term{X}, represents a set of node features extracted using the feature function \(f_{extract}\).
    The function \(f_{extract}\) is designed to extract and normalize the feature set based on the nodes \term{V}, edges \term{E}, and graph metadata \(\mathcal{K}\) of a provided transaction.
    Specifically, the function \(f_{extract}(\mathcal{V}, \mathcal{E}, \mathcal{K})\) accepts \(\mathcal{V}\), \(\mathcal{E}\) and \(\mathcal{K}\) as inputs and returns the feature set \(\mathcal{X}\).
    \item[Cash Flow Graph \term{G}:]
    The cash flow graph, denoted by \(\mathcal{G} = (\mathcal{V}, \mathcal{E}, \mathcal{X})\), represents the asset transfer information within a transaction.
\end{description}

\subsection{The Design of \tool{}}
%
In this paper, we introduce {\tool}, an automatic detection service designed to address the two aforementioned challenges and perform PMA detection.
To offer a high-level overview of {\tool}, we discuss the design purpose behind each component. 
As depicted in Figure~\ref{fig:design_overview}, {\tool} comprises three main components: {\it Transaction Parser}, {\it Graph Builder}, and {\it Graph Classifier}.

\noindent \textit{1)} The {\it Transaction Parser} component is designed to parse transactions collected from EVM-based blockchains to retrieve their transfer-related call traces and event logs.
As illustrated in Equation~\ref{eq:tp}, we define the function \(f_{parse}\) that accepts a transaction, \textit{tx}, as its input and returns the call traces and event logs as its output.
\begin{equation}
\label{eq:tp}
\begin{aligned}
    f_{parse}(\textit{tx}) = Traces_{\textit{tx}},\ Events_{\textit{tx}}
\end{aligned}
\end{equation}

\noindent \textit{2)} The {\it Graph Builder} is designed to construct a cash flow graph \(\mathcal{G}\) using the filtered call traces and event logs. 
Within this component, we also extract and normalize four distinct classes of node features to facilitate the learning process.
Specifically, the function \(f_{construct}\), as outlined in Equation~\ref{eq:gb_construction}, takes the filtered call traces and event logs as inputs and yields nodes \term{V}, edges \term{E}, and graph metadata \term{K} as outputs.
Additionally, the function \(f_{extract}\) in Equation~\ref{eq:gb_extraction} extracts the node features \term{X} and generates a cash flow graph \term{G} using the outputs from \(f_{construct}\).
\begin{equation}
\label{eq:gb_construction}
\begin{aligned}
    f_{construct}(Traces_{tx}, Events_{tx}) = (\mathcal{V}, \mathcal{E}, \mathcal{K})
\end{aligned}
\end{equation}

\begin{equation}
\label{eq:gb_extraction}
\begin{aligned}
    f_{extract}(\mathcal{V}, \mathcal{E}, \mathcal{K}) = \mathcal{G}
\end{aligned}
\end{equation} 

\noindent \textit{3)} The {\it Graph Classifier} is designed to predict whether the cash flow graph \term{G} corresponds to a PMA or non-PMA.
During the training phase, given a collection of cash flow graphs \(\mathcal{G}s\), associated labels \textit{Labels}, and a selected GNN algorithm \(\mathcal{M}\), the function \(f_{train}\) (refer to Equation~\ref{eq:gc_learning}) outputs a trained GNN model \(\mathcal{F}\).
For the inference phase, provided with a collection of cash flow graphs \(\mathcal{G}s\) and a trained GNN model \(\mathcal{F}\), the function \(f_{inference}\) (refer to Equation~\ref{eq:gc_inference}) generates a prediction result \term{P}.
This result indicates if the graph is a PMA (\term{P} = 1) or non-PMA (\term{P} = 0).

\begin{equation}
\label{eq:gc_learning}
\centering
\begin{aligned}
    f_{train}(\mathcal{G}s,\ Labels,\ \mathcal{M}) = \mathcal{F}
\end{aligned}
\end{equation} 

\begin{equation}
\label{eq:gc_inference}
\centering
\begin{aligned}
    f_{inference}(\mathcal{G}s, \mathcal{F}) = \mathcal{P},\ where\ \mathcal{P} = 0\ or\ 1 
\end{aligned}
\end{equation}

Further implementation details of each component in {\tool} can be found in Section~\ref{sec:implementation}.
\section{The Details of \tool{}}
\label{sec:implementation}

In this section, we delve deeper into the implementation details of \tool{}.

\begin{figure*}[ht]
    \centering
    \includegraphics[width=1\linewidth]{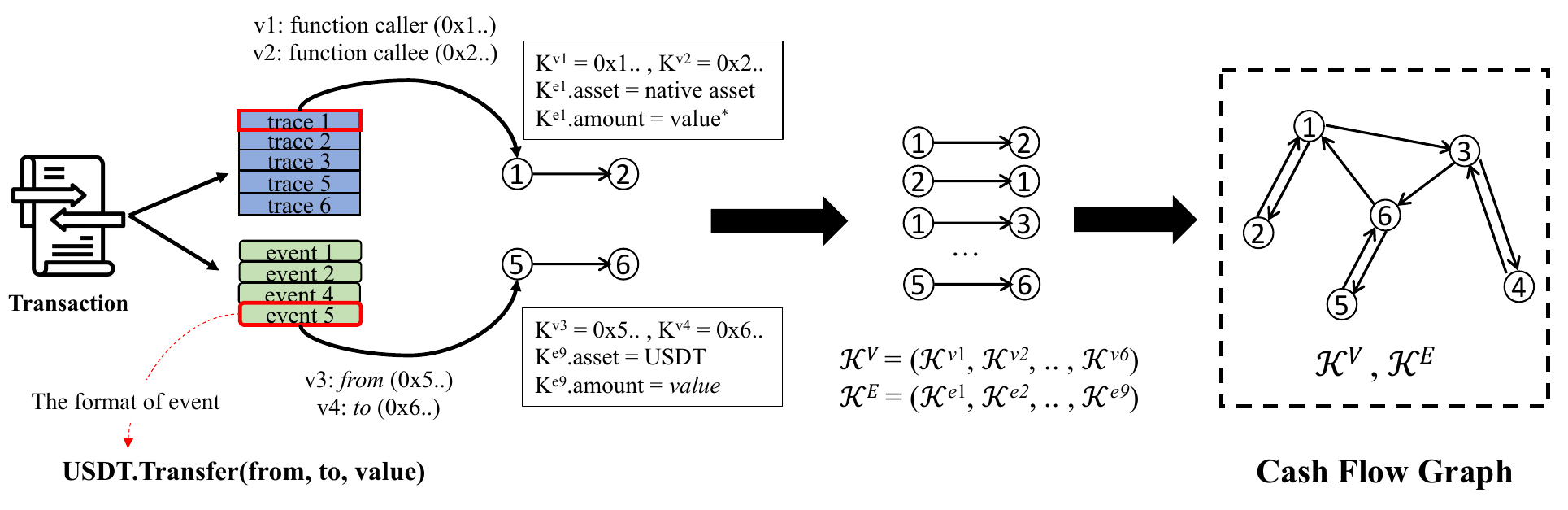}
    The `value*' represents the amount of native assets, which can be extracted from the value sector of the function call trace. 
    \caption{\bf The graph construction process.} 
    \label{fig:graph_construction}
\end{figure*}
\subsection{Transaction Parser}
\label{subsec:impl_collecctor}
The {\it Transaction Parser} aims to retrieve all transfer-related call traces and event logs from a transaction collected from the EVM-based blockchains.
In an EVM-based blockchain, a function call trace refers to a detailed record of the execution step taken during the invocation of a smart contract function.
As for the transaction event log, it is an immutable record pre-defined by the smart contract and produced during the transaction execution.

As shown in Figure~\ref{fig:design_overview}, three steps are followed to generate the transfer-related call traces and event logs for the subsequent component.
First, the {\it Transaction Parser} retrieves the raw transaction from EVM-based blockchains using a well-connected archive node.
Second, the {\it Transaction Parser} replays the raw transaction to capture all associated call traces and event logs.
Lastly, the {\it Transaction Parser} refines collected call traces and event logs by identifying the native and ERC20 standard asset transfers.
Specifically, the native token transfer can be identified from the call traces by examining their \code{value} sector.
Meanwhile, the ERC20 standard asset transfers are discerned based on the signature of the \code{Transfer} events.
It is worth noting that transactions without asset transfers are not forwarded to the next component.

After selecting all transfer-related call traces and event logs for a transaction, the {\it Transaction Parser} sends them to the subsequent component for graph construction.
Additionally, during the training phase, the transaction label is also provided to aid the training process.


\subsection{Graph Builder}
\label{subsec:impl_builder}
Upon receiving the filtered call traces and event logs from the {\it Transaction Parser}, the {\it Graph Builder} undertakes the tasks of graph construction and feature extraction, subsequently generating the cash flow graph.


\subsubsection{Graph Construction}
\label{subsubsec:construction}
As demonstrated in Figure~\ref{fig:graph_construction}, the {\it Graph Builder} iterates over all filtered call traces and event logs to translate them into a cash flow graph with the graph metadata. 
This translation is achieved by pinpointing senders, receivers, assets, and the respective asset amounts from both call traces and event logs.

%
The asset transfer regarding the native asset can be extracted from the call traces by analyzing three factors: caller, callee, and the sent native asset amount.
In a filtered call trace, the function caller acts as the asset sender (\(caller\) acts as a \(v_{sender}\)), while the function callee stands as the asset receiver (\(callee\) acts as a \(v_{receiver}\)).
Moreover, the caller and callee address will be recorded in \(\mathcal{K}^{V}\).
Subsequent to node identification, the edge \(e:\ v_{sender} \xrightarrow[]{\mathcal{K}^{e}} v_{receiver}\) is formulated, where \(\mathcal{K}^{e}\) encompasses both the asset type and its amount.
Specifically, the asset type refers to the native token (e.g., ether on the Ethereum blockchain), and the asset amount is derived from the \code{value} sector within the call trace.

%
The asset transfer regarding the ERC20 token can be extracted from the \code{Transfer} event.
As prescribed by the ERC20 token standard, the parameters of the \code{Transfer}~\footnote{Transfer(address indexed \_from, address indexed \_to, uint256 \_value)} event delineates the sender, receiver, and asset amount. The event emitter signifies the asset transferred.
After identifying the necessary information, the {\it Graph Builder} can produce the corresponding edge \(e\) and metadata \(\mathcal{K}^{v},\ \mathcal{K}^{e}\).
%

Once all nodes \term{V}, edges \term{E}, and their pertinent metadata \term{K} have been structured, the {\it Graph Builder} advances to extract features \term{X} to output the completed cash flow graph \term{G}.

\subsubsection{Feature Extraction}
\label{subsubsec:feat}
        

\begin{table*}[ht]
     \caption{\bf The node features.}
 	\label{tbl:feature}
	\centering
    \begin{tabular}{cccc}
		\toprule
		{\bf Feature} & {\bf Notation} & {\bf Description}& {\bf Value Range}\\
		\midrule
	      Node Type & $X_{type} $& The type and transparency of an account & $  [0,0,1] or [0,1,0] or [0,0,1]$\\
            \midrule
            Transfer Frequency & $X_{frequency}$ & The number of times an account sends/receives assets.  & $\in (0, 1]$\\
            \midrule
            Transfer Diversity & $X_{diversity}$ & The number of asset types sent/received by an account & $\in (0, 1]$\\
            \midrule
            Profit Score & $X_{profit}$ & The normalized value of the node's profit & $\in [-1, 1]$\\
        \bottomrule
 	\end{tabular}
        \newline
        
\end{table*}

Given the nodes \term{V}, edges \term{E}, and corresponding graph metadata \term{K}, the {\it Graph Builder} extracts four distinct classes (in Table~\ref{tbl:feature}) of node features for each node.


\noindent \textbf{Node Type.}
In the EVM-based blockchains, there are two types of addresses: EOAs and CAs.
Unlike EOAs, CAs can be triggered to execute complex logic based on their code.
Moreover, the majority of {\dapp}s disclose and authenticate their source code during deployment, thereby allowing users to scrutinize the intricacies of their business logic.
Conversely, smart contracts deployed by attackers typically manifest as unverified and lack transparency.
Therefore, the {\it Graph Builder} captures the feature {\it node type} by encapsulating both the account type and the transparency for each node. 

To verify the transparency of nodes, we constructed a key-value database, denoted as $DB_{account}$, through verifying pre-collected CA on {\it etherscan.io} and {\it bscscan.io}.
In the database, the key corresponds to the account address, while the value, a boolean value, indicates whether the collected CAs are verified with their source code.
Leveraging the $DB_{account}$ database, the {\it Graph Builder} employs the account address recorded in \(\mathcal{K}^{V}\) to retrieve both account type and associated transparency.

\begin{algorithm}[t]
\floatname{algorithm}{Feature Extraction}
\caption{Node Type}
\label{algo:node_type}
\begin{algorithmic} 
\REQUIRE $\mathcal{V},\ \mathcal{K}^{V},\ DB_{account}$
\ENSURE $\mathcal{X}_{type}$
\STATE $\mathcal{X}_{type} \gets \{\}$
\FOR{$(v, \mathcal{K}^{v})\ \in\ (\mathcal{V}, \mathcal{K}^{V})$}
    \STATE $account \gets \mathcal{K}^{v}$
    \IF{$DB_{account}[account]\ exists$}
        \IF{$DB_{account}[account]\ == \ False$}
            \STATE $\mathcal{X}_{type}[v] \gets [1,0,0]$   \texttt{ //Opaque CA}
        \ENDIF
        \STATE $\mathcal{X}_{type}[v] \gets [0,1,0]$   \texttt{ //Transparent CA}
    \ENDIF
    \STATE $\mathcal{X}_{type}[v] \gets [0,0,1]$    \texttt{ //EOA}
\ENDFOR
\RETURN $\mathcal{X}_{type}$
\end{algorithmic}
\end{algorithm}

As illustrated in Feature Extraction~\ref{algo:node_type}, the {\it Graph Builder} iterates \(\mathcal{V}\) and verifies the corresponding account address (retrieved from \(\mathcal{K}^{V}\)) in \(DB_{account}\).
As a result, the node type feature, denoted by \(\mathcal{X}_{type}\), is extracted for all nodes \term{V}.

\noindent \textbf{Transfer Frequency.}
An unusual number of asset transfers can act as a flag for potential price manipulation.
Repeatedly trading a pair of assets in a transaction is not prevalent and is mostly considered malicious behavior such as washing trading, price manipulation, or reentrancy attack.
Therefore, we capture the feature {\it transfer frequency} to represent a node's asset sending and receiving frequency.

As illustrated in Feature Extraction~\ref{algo:frequency}, given the nodes \term{V} and edges \term{E}, the {\it Graph Builder} iterates edges $\mathcal{E}$ and accumulates the number of incoming and outgoing edges for nodes \term{V}.
Then, the {\it Graph Builder} further normalizes each node's incoming and outgoing edge count by dividing the largest incoming and outgoing edge count.
As a result, the transfer frequency feature, denoted by \(\mathcal{X}_{freq}\) is extracted for nodes \term{V}.

\begin{algorithm}
\floatname{algorithm}{Feature Extraction}
\caption{Transfer Frequency}
\label{algo:frequency}
\begin{algorithmic} 
\REQUIRE $\mathcal{V},\ \mathcal{E}$
\ENSURE $\mathcal{X}_{frequency}$
\STATE $inEdge \gets \{\},\ outEdge \gets \{\},\ X_{freq} \gets \{\}$
\FOR{$\textit{e}\ \in\ \mathcal{E}$}
    \STATE $inEdge[e.receiver] ++$
    \STATE $outEdge[e.sender] ++$
\ENDFOR
\STATE $maxIn \gets max(inEdge)$ 
\STATE $maxOut \gets max(outEdge)$
\FOR{$\textit{v}\ \in\ \mathcal{V}$}
    \STATE $X_{frequency}[v] \gets [\frac{inEdge[v]}{maxIn},\ \frac{outEdge[v]}{maxOut}]$
\ENDFOR
\RETURN $\mathcal{X}_{frequency}$
\end{algorithmic}
\end{algorithm}

\noindent \textbf{Transfer Diversity.}
Sophisticated PMAs often aim to create artificial arbitrage opportunities such as the infamous bZx attack~\cite{bZxhack}. 
An unusual number of assets being involved in a transaction can be indicative of such an attack.
Therefore, we capture the feature {\it transfer diversity} to describe the number of different assets sent and received by the node.

As illustrated in Feature Extraction~\ref{algo:frequency}, given the nodes\term{V}, edges \term{E} and metadata \(\mathcal{K}^{E}\), the {\it Graph Builder} iterates edges $\mathcal{E}$ and accumulates the number of incoming and outgoing assets for nodes $\mathcal{V}$.
Then, {\it Graph Builder} further normalizes each node's incoming and outgoing asset count into the range of $(0,1]$ by dividing the largest incoming and outgoing asset count in the graph.
As a result, the transfer diversity feature, denoted by \(\mathcal{X}_{diversity}\) is extracted for nodes \term{V}.

\begin{algorithm}[tb]
\floatname{algorithm}{Feature Extraction}
\caption{Transfer Diversity}
\label{algo:diversity}
\begin{algorithmic} 
\REQUIRE $\mathcal{V},\ \mathcal{E},\ \mathcal{K}^{E}$
\ENSURE $\mathcal{X}_{diverisy}$
\STATE $inAsset \gets \{\},\ outAsset \gets \{\},\ X_{diverisy} \gets \{\}$
\FOR{$(\textit{e},\ \mathcal{K}^{e})\ \in\ (\mathcal{E}, \mathcal{K}^{E})$}
    \STATE $inAsset[e.receiver].append(\mathcal{K}^{e}.asset)$
    \STATE $outAsset[e.sender].append(\mathcal{K}^{e}.asset)$
\ENDFOR
\STATE $maxIn \gets MaxCount(inAsset)$ 
\STATE $maxOut \gets MaxCount(outAsset)$ 
\FOR{$\textit{v}\ \in\ \mathcal{V}$}
    \STATE $inCount \gets set(inAsset[v]).length$
    \STATE $outCount \gets set(outAsset[v]).length$
    \STATE $X_{diversity}[v] \gets [ \frac{inCount}{maxIn},\ \frac{outCount}{maxOut}]$
\ENDFOR
\RETURN $\mathcal{X}_{diversity}$
\end{algorithmic}
\end{algorithm}

\noindent \textbf{Profit Score.}
In a PMA, one or a few addresses often stand to gain disproportionately compared to others. 
Given the fact that the attacker's goal is to gain illicit profits, anomalously high profits for a particular node amidst a complex transaction can be a red flag pointing toward manipulation.
However, in a transaction, accurately calculating each node's actual profit requires obtaining the involved assets' decimals and the price data.
Acquiring such information is expensive and might cause time delays for the graph construction task due to third-party resource dependency.
Therefore, in this work, we propose a normalized value {\it profit score} to describe each node's profit regardless of the asset decimals and price data.

As demonstrated in Feature Extraction~\ref{algo:profit}, given the edges \term{E} and metadata \(\mathcal{K}^{E}\), the {\it Graph Builder} iterates metadata \(\mathcal{K}^{E}\) and finds the largest transfer amount for each asset.
Then, the {\it Graph Builder} iterates edges \(\mathcal{E}\) and accumulates the normalized amount for nodes based on the largest amount.
As a result, the normalized profit, denoted by \(\mathcal{X}_{profit}\), is calculated for all nodes \term{V}.

\begin{algorithm}
\floatname{algorithm}{Feature Extraction}
\caption{Profit Score}
\label{algo:profit}
\begin{algorithmic} 
\REQUIRE $\mathcal{E},\ \mathcal{K}^{E}$
\ENSURE $\mathcal{X}_{profit}$
\STATE $MaxAmount \gets \{\},\ \mathcal{X}_{profit} \gets \{\}$
\FOR{$\mathcal{K}^e\ \in\ \mathcal{K}^{E}$}
    \STATE $asset \gets \mathcal{K}^e).asset,\ amount \gets \mathcal{K}^e).amount$
    \IF{$amount > MaxAmount[asset]$}
        \STATE $MaxAmount[asset] \gets amount$
    \ENDIF
\ENDFOR

\FOR{$(e, \mathcal{K}^{e})\ \in\ (\mathcal{E}, \mathcal{K}^{E})$}
    \STATE $asset \gets \mathcal{K}^e.asset,\ amount \gets \mathcal{K}^e.amount$
    \STATE $\mathcal{X}_{profit}[e.sender] -= amount / MaxAmount[asset]$
    \STATE $\mathcal{X}_{profit}[e.receiver] += amount / MaxAmount[asset]$
\ENDFOR
\RETURN $\mathcal{X}_{profit}$
\end{algorithmic}
\end{algorithm}

In conclusion, after the feature extraction process, a set of node features, denoted by \(\mathcal{X} = X_{type} + X_{frequency} + X_{diversity} + X_{profit}\), is generated.
Subsequent to it, the {\it Graph Builder} feeds the completed cash flow graph \(\mathcal{G} = (\mathcal{V}, \mathcal{E}, \mathcal{X})\) to the {\it Graph Classifier} for the training and inference process.

\begin{figure*}[!ht]
    \centering
    \includegraphics[width=1\linewidth]{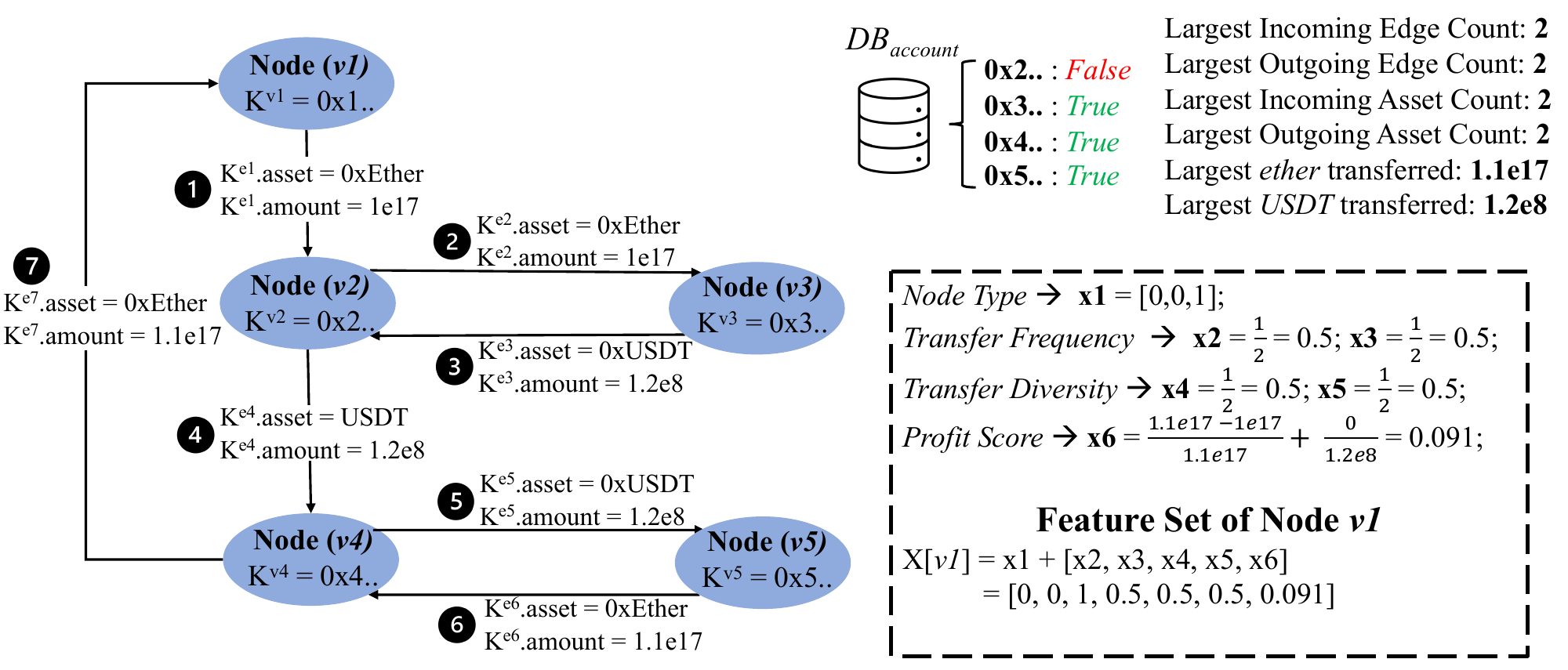}
    \caption{\bf An example of a cash flow graph with four distinct features.} 
    \label{fig:5_illustrative_example}
\end{figure*}

\subsubsection{An Illustrative Example}
To help understand the features, we elaborate on an illustrative example in Figure~\ref{fig:5_illustrative_example}.
Taking the node $v1$ as an example, since the account address of node $v1$ (i.e., the value of \(\mathcal{K}^{v1}\)) does not exist in the \(DB_{account}\), so the \(x1\) of \(v1\) is $[0,0,1]$.
The node $v1$ has 1 incoming and 1 outgoing edges. Based on analyzing the cash flow graph, we discover that the largest incoming and outgoing edge counts are both 2.
Therefore, we can further calculate the value of $x2 = \frac{1}{2} = 0.5$ and $x3 = \frac{1}{2} = 0.5$ representing node $v1$'s transfer frequency.
Similary, we can also calculate the value of $x3 = \frac{1}{2} = 0.5$ and $x4 = \frac{1}{2} = 0.5$ representing node $v1$'s transfer diversity.
As for the {\it profit score} of the node $v1$, there are only two assets circulated in this example and the node $v1$ sent $1e17 ether$ and received $1.1e17 ether$.
Thus, we can calculate $v1$'s profit score $x6 = \frac{1.1e17 - 1e17}{1.1e17} + \frac{0}{1.2e8} \approx 0.091$.
As a result, given all values $x1$, $x2$, $x3$, $x4$, $x5$, and $x6$, we formulate the feature vector for the node $v1$, as denoted \(\mathcal{X}[v1] = x1 + [x2, x3, x4, x5, x6] = [0, 0, 1, 0.5, 0.5, 0.5, 0.5, 0.091]\).

\subsection{Graph Classifier}
\label{subsec:impl_classifier}
As the last component of {\tool}, the {\it Graph Classifier} will be developed for predicting whether the transaction is benign or malicious based on the outputs from the {\it Graph Builder}. 
This component consists of two phases, the training phase which learns a GNN model, and the inference phase which predicts the labels based on the well-trained GNN. 

\noindent \textbf{Training Phase.}
During the training phase, the {\it Graph Classifier} component trains the GNN model with cash flow graphs \term{G} and their corresponding labels fed by the {\it Graph Builder} component. 
Then, it feeds the pre-trained GNN models to the next component for prediction. 
To fully utilize the fact that a raw transaction can be constructed as a cash flow graph, we intuitively choose to build a GNN model that directly learns graph knowledge and predicts graphs' labels.
In this work, we set out to explore multiple edge-cutting GNN algorithms that are able to conduct graph representation learning by incorporating neighbor node features and graph structural features for the graph classification task.
Therefore, we can directly apply underlying GNN algorithms and output the trained model.
Specifically, we set up to have two layers (i.e., {\it L}=2) using the GNN algorithms in DGL~\cite{wang2019deep}. 

\noindent \textbf{Inference Phase.}
During the inference phase, the {\it Graph Classifier} loads the pre-trained GNN model from the {\it GNN learner} component and predicts the labels of new cash flow graphs \term{G} fed by the {\it Graph Builder}.
Particularly, the {\it Graph Classifier} classifies \term{G} based on its returned binary numbers `0' and `1' indicating {\it benign} and {\it malicious} transaction respectively.
\section{Dataset}
\label{sec:data}
In this section, we first present the basic statistics of the collected dataset.
Then, we conduct a distribution analysis in terms of transaction complexity. 

\subsection{Dataset Collection}
We collected 208 PMA and 2,080 non-PMA transactions in our transaction collection, sampled from the time range between Jan 2021 and Jan 2022.
The PMA transactions were painstakingly curated by analyzing over 100 incidents reported in a public source~\footnote{https://de.fi/rekt-database}. 
Meanwhile, for the non-PMA transactions, we adopted a random sampling approach, targeting transactions on both Ethereum and BSC blockchains.

To maintain a high-quality dataset, we excluded transactions devoid of any asset transfers. 
Furthermore, to better mirror real-world complexities, we categorized our non-PMA transactions based on the number of asset transfers they encompassed. 
In particular, our complex non-PMA transaction subset only features transactions with more than 100 asset transfers.
Table~\ref{tbl:data_dist} reveals that the volume of non-PMA transactions we gathered is tenfold the PMA transactions.
Specifically, the collection of non-PMA transactions consists of an even split: 1,040 simple and 1,040 complex transactions.

\subsection{Distribution Analysis}
\label{subsec:tx_dist}
\begin{figure}[!ht]
    \centering
    \includegraphics[width=1\linewidth]{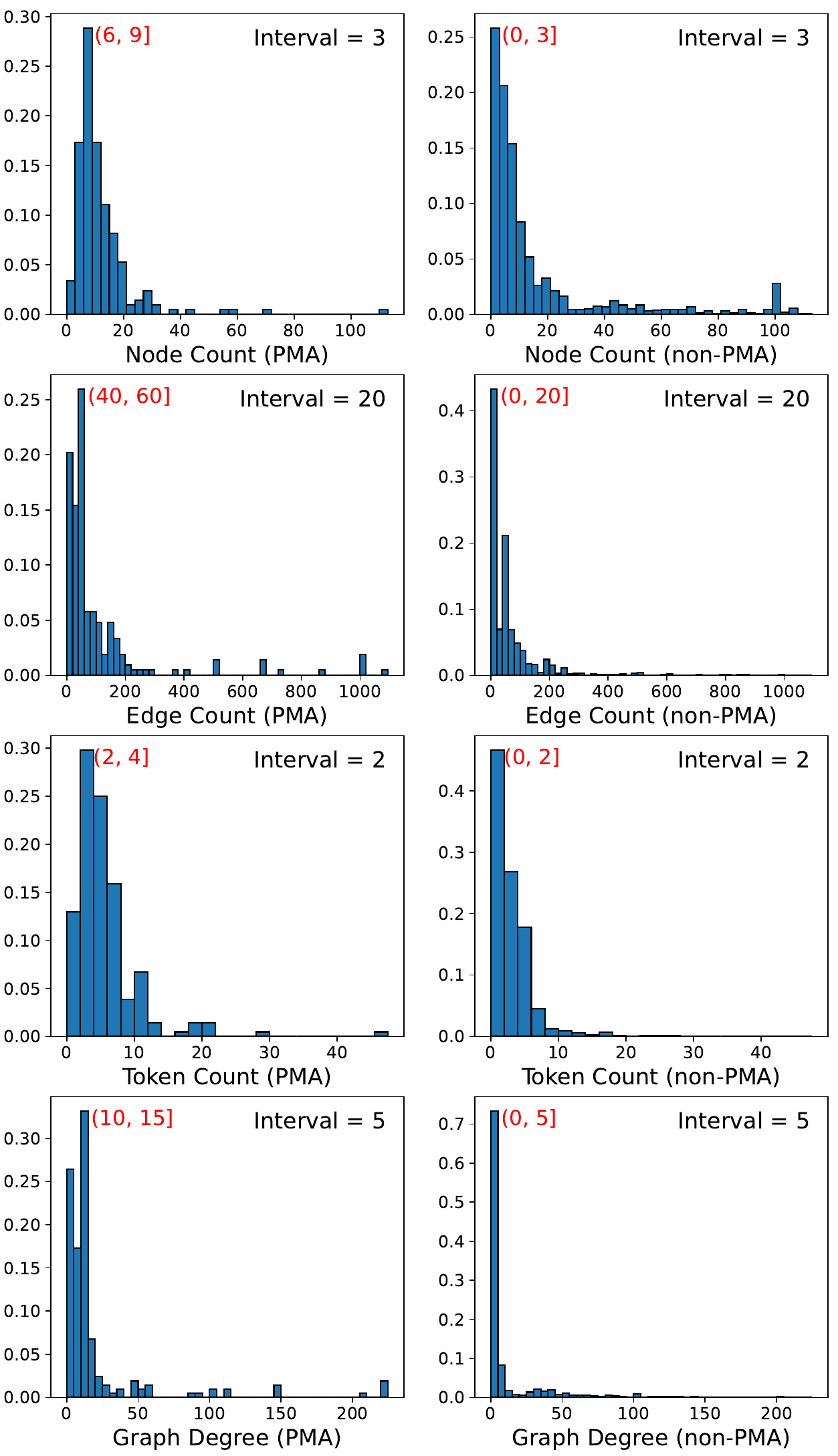}
    \caption{\bf Probability distribution analysis for graph-based metrics.}
    \label{fig:pdfs_general}
\end{figure}

\begin{figure}[t]
    \centering
    \includegraphics[width=1\linewidth]{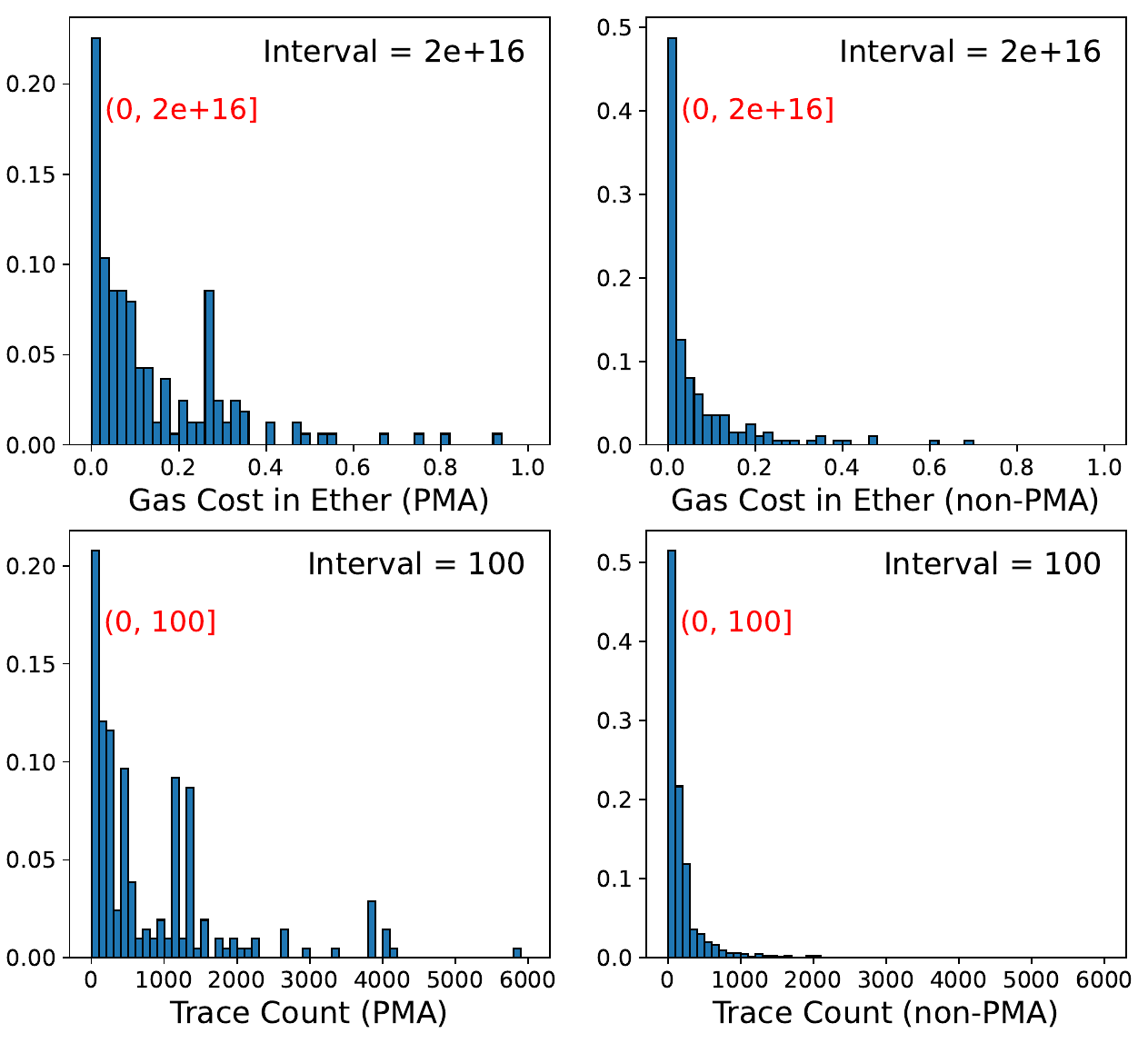}
    \caption{\bf Probability distribution analysis for transaction-based metrics.} 
    \label{fig:pdfs_tx}
\end{figure}

We present the basic statistics of collected transactions in terms of graph-based and transaction-based metrics in Table~\ref{tbl:data_dist}.
Moreover, we plot probability distribution figures for each metric to show the difference between collected PMA and non-PMA transactions in Figure~\ref{fig:pdfs_general} and \ref{fig:pdfs_tx}.

\begin{table}[h!]
    \caption{\bf The statistics of collected transactions in terms of graph- and transaction-based metrics.}
    \label{tbl:data_dist}
    \centering
    \resizebox{1\linewidth}{!}{
    \begin{tabular}{p{1.8cm}<{\centering}ccccc}
        \toprule
        \multirow{2}{*}{\bf Aspects} & \multirow{2}{*}{\bf Metrics} & \multirow{2}{*}{ \bf PMA} & \multicolumn{3}{c}{\bf non-PMA}\\
        & & & overall& simple & complex \\
        \midrule
        \multirow{4}{*}{Graph-based}&Node Count & 12.4 & 38.8 & 6.9 & 70.7  \\
        &Edge Count & 114.3 & 73.7 & 11.1 & 136.4  \\
        &Asset Count & 6.0 & 3.5 & 3.5 & 3.5  \\
        &Node Degree*& 21.5 & 18.9 & 3.1 &34.7  \\
        \midrule
        \multirow{2}{*}{Tx-based}&Gas Cost (Ether) & 1.4429 & 0.0715 & 0.0367  & 0.1063 \\
        &Trace Count & 858.2 & 192.7 & 73.5  & 311.9 \\
        \midrule
        \multicolumn{2}{c}{Transaction Count} & 208 & 2,080 & 1,040 & 1,040 \\
        \bottomrule
    \end{tabular}
    }
\end{table}

\noindent {\bf Graph-based Analysis.}
Upon analyzing the dataset using graph-based metrics such as node count, edge count, asset count, and node degree, we observed that PMA transactions have higher average values for edge count, asset count, and node degree compared to non-PMA transactions. 
Probability distribution analysis further revealed distinct patterns: PMA transactions show peak occurrences in the ranges of (6,9] for node count, (40,60] for edge count, (2, 4] for asset count, and (10,15] for node degree. 
Conversely, non-PMA transactions fall within the initial range for these metrics, highlighting a marked differentiation between the two transaction types.

\noindent {\bf Transaction-based Analysis.}
Upon scrutinizing transaction complexities using the transaction-based metrics (i.e., gas cost and trace count), distinct differences emerge between PMA and non-PMA transactions.
The statistical result (in Table~\ref{tbl:data_dist}) indicates that PMA transactions incur an average gas cost of 1.4429 ether, in contrast to a significantly lower 0.0715 ether for non-PMA transactions. 
Additionally, the trace count for PMA transactions stands at a substantial 858.2 on average, surpassing the non-PMA average of 192.7. 
The probability distribution of these metrics further clarifies this distinction. 
About 50\% of non-PMA transactions have a gas cost confined to the range [0, 2e+16], whereas a mere 25\% of PMA transactions adhere to this interval. 
Instead, the latter's gas cost displays a conspicuous spike within the range (2.6e17, 2.8e17]. 
In terms of trace count, while over half of the non-PMA transactions fall within the bounds of (0, 100], a mere 20\% of PMA transactions do so, with notable peaks observed in the intervals (1100,1200] and (1300,1400]. 
Notably, the probability distribution for non-PMA transactions reveals an exponential decline as both gas cost and trace count increase.

In conclusion, the collected PMA transactions are more complex than the sampled non-PMA transactions in terms of graph-based and transaction-based metrics.

\section{Evaluation}
\label{sec:evaluation}
In this section, we conduct to evaluate \tool{} by answering the following research questions:
\begin{itemize}
    \item \textbf{RQ1}: How does the proposed graph-based approach perform in detecting PMAs across various graph neural network models?
    \item \textbf{RQ2}: How significantly do extracted features enhance the detection of PMAs?
    \item \textbf{RQ3}: How practical is our service in identifying PMA transactions within real-world blockchains?
\end{itemize}

To answer \textbf{RQ1}, we assess the performance of the PMA detection task using several leading GNN models, including GCN, GAT, GraphSage, and GIN, in comparison with a baseline model, MLP. For a concise evaluation of the models' effectiveness, we consider a range of metrics, including Accuracy, TPR, FPR, and AUC-ROC, for our trained classifiers.

To answer \textbf{RQ2}, we undertake an ablation study, highlighting the impact of our proposed features on the PMA detection task. 
We evaluate several variants, namely: \textit{without node type}, \textit{without transfer frequency}, \textit{without transfer diversity}, and \textit{without profit score}.

To answer \textbf{RQ3}, we measure the time cost of each component in \tool{} in the real-world scenario.
Specifically, we start the evaluation from the phase of collecting newly processed transactions on the chain to the phase of predicting whether the transaction is launching the PMA.

\subsection{Experimental Setup}

\noindent \textbf{Selected Models.}
For the assessment of GNNs in detecting PMAs, we have chosen four cutting-edge GNN models (i.e., GCN~\cite{kipf2016semi}, GAT~\cite{velickovic2017graph}, GraphSAGE~\cite{hamilton2017inductive}, GIN~\cite{xu2018powerful}) and a baseline model (i.e., MLP~\cite{defferrard2016convolutional}).
%

\noindent \textbf{Hyperparameters.}
For all models presented in Section~\ref{subsec:eval_perf} and Section~\ref{subsec:eval_ablation}, we used ReLU as the non-linear activation function and Adam optimizer as the optimization algorithm.
All GNN and the baseline models we devised consist of two layers, including a hidden layer with a dimensionality of 16 and an output layer with a dimension of 2 to conduct a binary classification task.
Based on the experiments on evaluating the hyperparameters in Section~\ref{subsec:eval_perf}, we trained all models for 100 epochs with a balanced train size of 100.

\noindent \textbf{Hardware and Software.}
We conduct our evaluations on a Mac Mini machine, equipped with an Apple M1 chip (8-core CPU) and 16GB of RAM.
Besides, \tool{} is implemented with Python 3.8.9, Pytorch 1.3.1~\footnote{https://github.com/pytorch/pytorch}, and DGL 0.9.1~\footnote{https://github.com/dmlc/dgl}.

\noindent \textbf{Evaluation Metrics.}
In evaluating the performance of binary classification, we employ \textit{Accuracy}, \textit{TPR}, \textit{FPR}, and \textit{AUC-ROC} metrics that are recognized and utilized in assessing binary classification outcomes:
%
%
1) Accuracy is a metric used to measure the percentage of correctly predicted outcomes;
2) True positive rate (\(TPR = \frac{TP}{ TP + FN}\)) is a metric used to evaluate the ability to correctly identify positive cases in the data;
3) False positive rate (\(FPR = \frac{FP}{FP + TN}\)) is a metric used to evaluate the ability to falsely identify negative cases as positive;
4) Area under the curve (AUC-ROC, as shown in Equation~\ref{eq:auc}) is a common metric used to evaluate the performance of binary classification models. 
%
\begin{equation}
\label{eq:auc}
\begin{aligned}
\text{AUC} = \int_{-\infty}^{\infty} TPR\left(FPR^{-1}(x)\right) , dF(x)
\end{aligned}
\end{equation} 

\subsection{Performance Analysis (RQ1)}
\label{subsec:eval_perf}
\noindent \textbf{Overall Performance.}
We systematically assessed the performance of four GNN models against a baseline model using our collected dataset.
A concise summary of their performance, given identical training and testing configurations, is delineated in Table~\ref{tbl:performance}.
Evidently, \tool{} when integrated with GNN models consistently outperformed the baseline across all evaluated metrics.
For instance, the combination of \tool{} with the GraphSage model emerges as particularly distinguished, achieving 93.25\% in Accuracy, 91.67\% in TPR, 6.67\% in FPR, and 96.18\% in AUC. 
From the analysis, several key insights can be gleaned. 
Primarily, \tool{}, when parameterized with a train size of 100, delivers respectful performance spanning \textit{Accuracy}, \textit{TPR}, \textit{FPR}, and \textit{AUC-ROC}.
This validates the effectiveness of our cash flow graph design and four introduced features.
On the other hand, when compared to the baseline \textit{MLP} model, all GNN models exhibited enhanced performance. 
This indicates \tool{}, when paired with GNN models, effectively utilizes the proposed features to identify structural variances, ensuring accurate classification of PMAs.


\begin{table}[ht]
    \caption{\textbf{Summary of the models' performance for PMA detection on our collected dataset.} }
 	\label{tbl:performance}
	\centering
    \begin{tabular}{c|p{1cm}<{\centering}p{1cm}<{\centering}p{1cm}<{\centering}p{1cm}<{\centering}}
            \toprule
		Model & Accuracy & TPR & FPR & AUC\\
            \midrule
            MLP (baseline) & 0.8223 & 0.8333 & 0.1783 & 0.8911 \\
            GCN & 0.8530 & 0.8796 & 0.1485 & 0.9341 \\
            GAT & 0.9009 & 0.8704 & 0.0975 & 0.9518 \\
            GIN & 0.8812 & 0.9167 & 0.1207 & 0.9544\\
            GraphSAGE & \textbf{0.9325} & \textbf{0.9167} & \textbf{0.0667} & \textbf{0.9618} \\
        \bottomrule
 	\end{tabular}
\end{table}

\noindent \textbf{Tuning Epoch\&Train Size.}
\begin{figure*}[ht]
    \centering
    \includegraphics[width=1\linewidth]{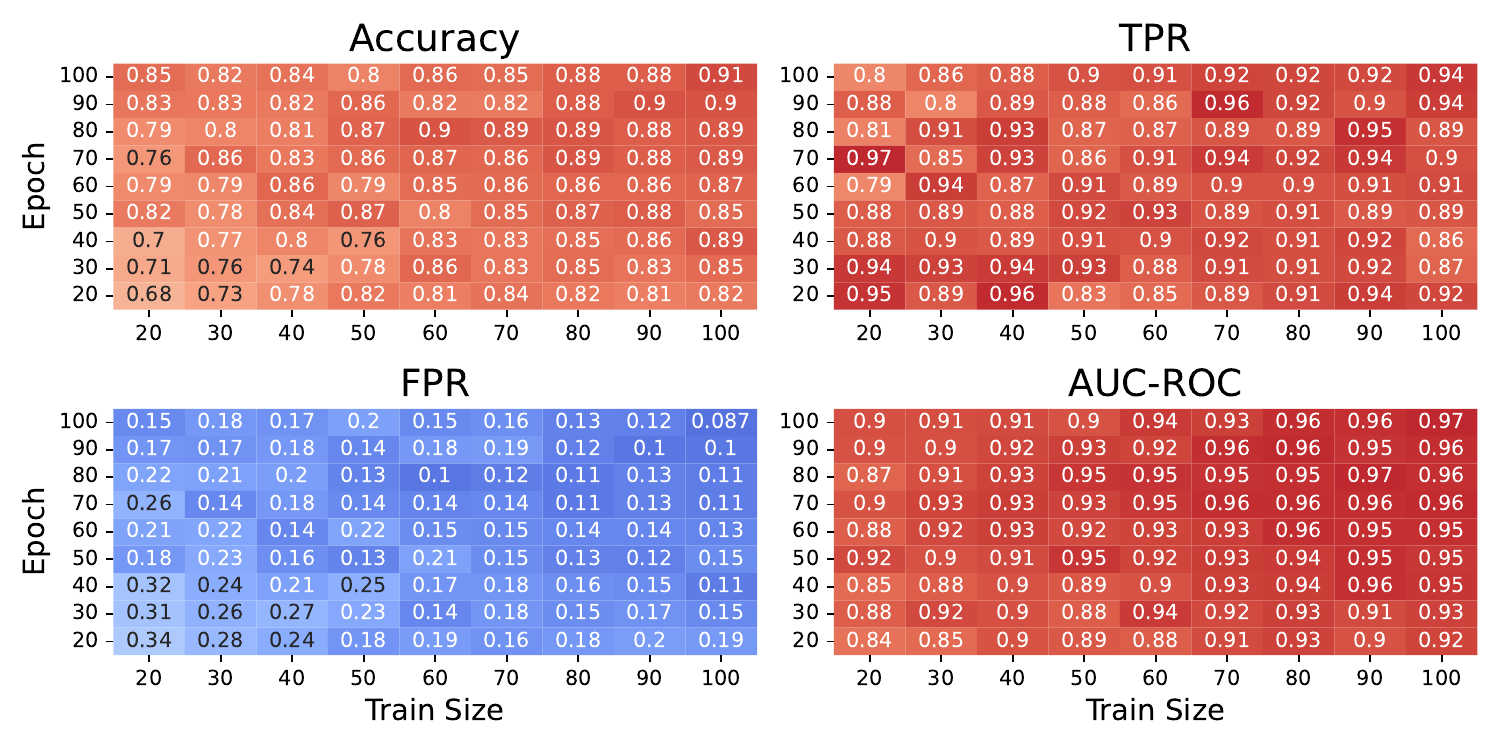}
    \vspace{-20pt}
    \caption{\bf The summary of GraphSage model's performance in the hyperparameter tuning process.} 
    \vspace{-10pt}
    \label{fig:perf_graphsage}
\end{figure*}
In the optimization process of our model, it is imperative to determine the most suitable hyperparameters to ensure both robustness and efficiency. 
In the tuning process, two critical parameters \textit{Epoch} and \textit{Train Size} are under consideration.
The epoch determines how many times the learning algorithm will work through the entire training dataset, while the training size indicates the number of PMA and non-PMA transactions used for each learning iteration.
To pinpoint the optimal pairing of these hyperparameters, we systematically evaluated various combinations and monitored their respective performances.

Using the GraphSage algorithm as a representative example, we visualized the results through a series of heatmaps (in Figure~\ref{fig:perf_graphsage}), each emblematic of a specific performance metric: Accuracy, TPR, FPR, and AUC-ROC. Each heatmap delineates train size on the x-axis and epoch on the y-axis. The epoch varies from 20 to 100 with intervals of 10, and analogously, the train size ranges similarly. 
Our observations from this evaluation are multifaceted. 
Firstly, the classifier's performance in terms of Accuracy, FPR, and AUC-ROC escalates with increasing values of both the epoch and train size. 
Secondly, it's discernible that the classifier can yield respectable accuracy even with relatively smaller train sizes and epochs, specifically greater than 60 in both dimensions.
Most notably, the classifier manifests its peak performance in Accuracy, FPR, and AUC-ROC metrics at the right top of the heatmaps (epoch = 100 and train size = 100), while its TPR is the second best.
Given this superior performance profile, we have elected to adopt the hyperparameters of \code{epoch = 100} and train \code{size = 100} for our subsequent evaluations.

\begin{framed}
    \noindent \textbf{Answer to RQ 1:} \textit{
    Upon integration with GNN models, \tool{} markedly outperforms the baseline \textit{MLP} model. 
    The distinct advantage of GNN models lies in their ability to recognize structural differences, an aspect that the MLP model appears to be less efficient regarding PMA detection.
    Moreover, the tuning results suggest that training with only 100 PMA and non-PMA transactions across 100 epochs yields strong performance, highlighting the cash flow graph's effectiveness in capturing the trading behavior of the transaction.
    }
\end{framed}

\subsection{Ablation Study (RQ2)}
\label{subsec:eval_ablation}

\begin{figure}[ht]
    \centering
    \includegraphics[width=1\linewidth]{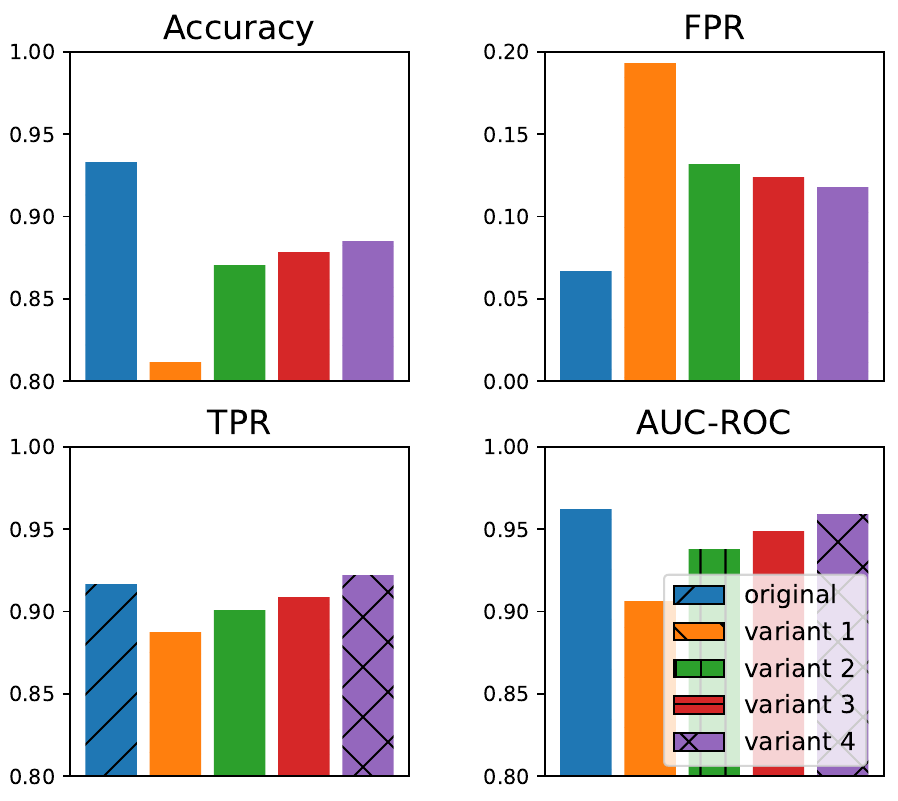}
    \caption{\bf The ablation study on four variants.} 
    \label{fig:ablation}
\end{figure}

To investigate how node features embedded in the cash flow graph impact the performance of \tool{}, we conducted ablation studies on four variants: 
\textit{1) variant without node type}, 
\textit{2) variant without transfer frequency},
\textit{3) variant without transfer diversity},
and \textit{4) variant without the profit score}. 

Using the GraphSage model as an illustrative example, Figure~\ref{fig:ablation} depicts the performance metrics for the four variants in relation to the original, which integrates all extracted features. 
Collectively, the performance of these variants of \tool{} is inferior to the original.
%
%
When the node type is excluded (\textit{variant 1}), there is a decrease in \tool{}'s \textit{Accuracy} by 12.10\%, \textit{TPR} by 2.96\%, and \textit{AUC-ROC} by 5.54\%. Conversely, its \textit{FPR} escalates by 12.59\%.
%
%
In the absence of the transfer frequency feature (\textit{variant 2}), the \textit{Accuracy} of \tool{} diminishes by 6.22\%, \textit{TPR} by 1.57\%, and \textit{AUC-ROC} by 2.39\%. Meanwhile, its \textit{FPR} rises by 6.47\%.
%
%
Omitting the transfer diversity feature (\textit{variant 3}) results in a decline in \tool{}'s \textit{Accuracy} by 5.43\%, \textit{TPR} by 0.83\%, and \textit{AUC-ROC} by 1.32\%. Additionally, there's an increment in \textit{FPR} by 5.67\%.
%
%
Without the profit score (\textit{variant 4}), \tool{} witnesses a reduction in \textit{Accuracy} and \textit{AUC-ROC} by 4.80\% and 0.28\% respectively. On the other hand, both \textit{TPR} and \textit{FPR} exhibit increases of 0.55\% and 5.08\% respectively.
The result implies that \tool{} is contributed by all features rather than certain dominating factors.

\begin{framed}
    \noindent \textbf{Answer to RQ 2:} \textit{
    The ablation results indicate that the efficacy of \tool{} is contributed by the combination of all extracted features.
    This collaborative feature interplay ensures that \tool{}'s performance is not dominated by any single factor.
    }
\end{framed}

\subsection{Time Cost (RQ3)}
\label{subsec:eval_timecost}

Detecting and responding to attacking transactions in a timely manner is crucial to minimize the potential damage caused by malicious activities in the \defi ecosystem.
A service with low time cost can quickly identify suspicious transactions, trigger alerts, and enable swift mitigation measures to prevent further harm to the \defi ecosystem.
To assess the practicality of \tool{}, we conduct a comprehensive evaluation by quantifying the time cost associated with detecting PMA transactions.
This evaluation encompasses the entire workflow, encompassing the transactions parsing, graph construction, and graph classification.
By measuring the time required for each stage, we can effectively evaluate the efficiency and feasibility of \tool{} in detecting PMAs (in Table~\ref{tbl:time_cost}).

\begin{table}[h!]
    \caption{\bf Time cost.}
    \label{tbl:time_cost}
    \centering
    \resizebox{1\linewidth}{!}{
    \begin{tabular}{cccc}
        \toprule
        \multirow{2}{*}{\bf Phase} & \multirow{2}{*}{\bf Malicious} & \multicolumn{2}{c}{\bf Benign}\\
        & & complex & simple \\
        \midrule
        Transaction Parsing & 4,383ms & 1,198ms & 394ms \\
        Graph Construction & 934ms & 894ms  & 435ms \\
        Graph Classification &  0.093ms & 0.098ms & 0.090ms \\
        \midrule
        Total Time Cost & 5,317ms & 2,047ms & 829ms \\
        \bottomrule
    \end{tabular}
    }
\end{table}

\noindent \textbf{Transaction Parsing.}
In the transaction parsing phase, we measure the time cost spent by the \textit{transaction parser} in extracting call traces and event logs from the raw transaction.
The time consumption in this phase depends on the amount of original call traces and event logs retrieved after the transaction replaying process.
As a result, embedded with a node well-connected to the blockchain network, \textit{transaction parser} spends 4,383ms, 1,198ms, and 394ms to extract call traces and event logs from the PMA, complex non-PMA, and simple non-PMA transactions on average.
The reason why parsing the PMA transaction has the highest time cost is that the PMA transaction normally consists of more invocation to execute the complex attacking logic.


\noindent \textbf{Graph Construction.}
In the graph construction phase, we measure the time cost spent by the \textit{graph builder} in constructing the cash flow graph and extracting the corresponding features.
The time consumption in this phase depends on the amount of call traces and event logs fed by the {\it transaction parser}.
As a result, \textit{graph builder} spends 934ms, 894ms, and 435ms on average to extract call traces and event logs from the PMA, complex non-PMA, and simple non-PMA transactions.

\noindent \textbf{Graph Classification.}
In the graph classification phase, we measure the time cost spent by \textit{Graph Classifier} in predicting the received cash flow graphs.
As a result, the {\it graph classifier} consumes less than 0.1ms (0.094ms for a single transaction on average) to predict all types of transactions and provides an average throughput of  10,605 transactions per second.

In conclusion, \tool{} spends  5,317ms, 2,047ms, and 829ms to complete the classification task from the parsing to predicting. 
It is worth noting that the time cost of predicting all types of transactions is less than the block mining time (12-14 seconds) on Ethereum.
With this rapid classification, the project can deploy \tool{} to evaluate the transaction interacting with their smart contracts.
The rapid detection enables the project to activate the pausing mechanism to prevent potential loss.

\begin{framed}
    \noindent \textbf{Answer to RQ 3:} \textit{
        Overall, the time cost of completing the classification for a single transaction ranges from 0.892 to 5.317 seconds, which is less than the time of creating a new block on Ethereum.
        The result implies that \tool{} is feasible and practical in terms of time cost so that the victims (including {\dapp}s and users) can have sufficient time to rescue their vulnerable assets. 
    }
\end{framed}




\section{Related Work}
\label{sec:related}

\noindent \textbf{DeFi Security.}
Smart contract vulnerability detection is crucial for ensuring the security and reliability of the DeFi ecosystem.
Given the irreversible and transparent nature of blockchain transactions, smart contract vulnerabilities can result in significant financial losses. 
Numerous academic endeavors have delved into this domain, leveraging a spectrum of methods spanning static analysis, dynamic analysis, and learning-driven methods. 
Static analysis identifies vulnerabilities through systematic code path probing (e.g., symbolic execution~\cite{luu2016making, albert2018ethir, nikolic2018finding, torres2019art, mossberg2019manticore, wang2019detecting, torres2018osiris, bose2022sailfish, so2021smartest}) and pattern recognition (i.e., formal verification~\cite{azzopardi2018monitoring, frank2020ethbmc}). 
Dynamic analysis (e.g., fuzzing techniques~\cite{schneidewind2020ethor, brent2020ethainter, ghaleb2022etainter, kalra2018zeus, tsankov2018securify, contro2021ethersolve, grech2018madmax, rodler2018sereum, feist2019slither, tikhomirov2018smartcheck}) examines the hidden vulnerabilities in in-execution analysis.
Meanwhile, learning-based methods have also shown great promise in detecting vulnerabilities (e.g., utilizing GNNs for scrutinizing control- and data-flow graphs extracted from smart contracts~\cite{zhuang2021smart, liu2021smart, liu2021combining}). 
Beyond the realm of smart contract vulnerability detection, a substantial body of research is probing security concerns associated with token systems, {\dapp}s, and other related facets. Qin et al.~\cite{qin2022quantifying, qin2023blockchain} investigated extractable values latent in the blockchain network, subsequently introducing an attack strategy that emulates profitable transactions sourced from the P2P network. 
The scholarly discourse has extensively addressed security predicaments, including but not limited to, attack detection~\cite{gai2023blockchain}~\cite{wu2021defiranger}, front-running~\cite{eskandari2019sok,daian2020flash,zhou2020highfrequency}, governance issues~\cite{gudgeon2020decentralized}, flash loan attack~\cite{qin2020attacking}. Complementing these, Sam et al.~\cite{werner2021sok} executed a comprehensive assessment of security challenges, both theorized and those manifesting in real-world scenarios.

\noindent \textbf{GNNs for Cybersecurity.}
GNNs have emerged as a powerful tool in the field of cybersecurity due to their ability to model complex relationships between entities and events. 
Particularly, GNNs can effectively capture the structural dependencies and interactions to detect and mitigate cyber threats in the field of code vulnerability detection~\cite{MirskyMBYPDML23, zhou2019devign, cheng2021deepwukong, wang2020combining, nguyen2022regvd}, network intrusion~\cite{yang2022wtagraph}, and spam detection in social networks~\cite{wang2019mcne, fan2019graph, bian2020rumor}.
For instance, Mirsky et al.~\cite{MirskyMBYPDML23}. introduced VulChecker, a tool employing a new program representation, slicing strategy, and message-passing graph neural network that precisely pinpoints and classifies vulnerabilities in source code.
King et al.~\cite{kingeuler} proposed a framework combining GNNs and recurrent neural networks and achieving state-of-the-art performance in anomalous lateral movement detection.
Yang et al.~\cite{yang2022wtagraph} designed a web tracking and advertising detection framework based on GNNs to analyze HTTP network traffic. 
Bian et al.~\cite{bian2020rumor} introduced the Bi-Directional Graph Convolutional Networks (Bi-GCN), a novel bi-directional graph model that captures both propagation and dispersion characteristics of rumors on social media.
\section{Conclusion}
\label{sec:conclusion}
This paper introduces the novel detection service, \tool{}, which utilizes Graph Neural Networks (GNNs) for PMA detection. 
By transforming raw transactions into cash flow graphs enriched with four distinct node features and capitalizing on the advantages of GNN models, \tool{} consistently surpasses traditional models across various metrics. 
Furthermore, time cost evaluations validate \tool{}'s efficiency, ensuring potential victims have sufficient time to secure their assets upon PMA detection. 
This work serves as a pivotal advancement in safeguarding the DeFi landscape from PMAs.

\bibliographystyle{IEEEtran}
\bibliography{main}

\clearpage

\end{document}